\documentstyle[pra,aps,epsfig]{revtex}

\begin{document}
\author{Giacomo Ciaramicoli, Paolo Tombesi and David Vitali}
\title{Performance of a deterministic source of entangled photonic qubits}
\address{Dipartimento di Matematica e Fisica, Universit\`a di
Camerino, \\
INFM, Unit\`a di Camerino, via Madonna delle Carceri 62032, Camerino, Italy \\}
\maketitle
\begin{abstract}
We study the possible limitations and sources of 
decoherence in the scheme for the deterministic generation of 
polarization-entangled photons, recently proposed by Gheri {\it et al}. 
[K. M. Gheri {\it et al}., Phys. Rev. A {\bf 58}, R2627 (1998)], 
based on an appropriately driven single atom trapped within an optical 
cavity.  
We consider in particular the effects of laser intensity fluctuations,
photon losses, and atomic motion.
\end{abstract}


\section{Introduction}

The efficient implementation of most quantum communication protocols
needs a controlled source of entangled qubits. Presently, the 
most common choice is using polarization-entangled photons, since they are able to
freely propagate maintaining the quantum coherence
between the different polarization components for long distances.
Significant demonstrations of this fact are the recent achievements
in experimental
quantum cryptography \cite{cryp} and quantum teleportation \cite{telep}.

The presently used source of polarization-entangled photons is parametric 
downconversion 
in $\chi^{(2)}$ nonlinear crystals (see, for example, \cite{kwiat}),
which however presents many disadvantages. The entangled
photons are in fact generated at random times, with a very low efficiency, and 
many photon properties are largely untailorable. Moreover,
the number of entangled qubits that can be produced directly with 
down-conversion is intrinsically limited. In fact, even if
a maximally entangled state of three photons, the so-called GHZ state \cite{ghz},
has been recently conditionally generated using two pairs of twin photons and 
the detection of one of them \cite{ghz2}, in this cases one generally 
needs higher order nonlinear processes, which are however extremely 
inefficient.
For this reason, the search for new photonic sources, able to generate
single-photon wave packets on demand, as for example {\em photon guns}
\cite{law}, or {\em turnstile devices} \cite{yama}, either entangled 
or not, is still very active.

Recently, Gheri {\it et al}. have proposed a controlled source of
entangled photons based on a cavity-QED scheme 
\cite{gheri}, which essentially generalizes the Law and Kimble photon pistol
scheme of Ref.~\cite{law} in such a way as to be able to produce 
polarization-entangled states of temporally separated single-photon
wave packets. This proposal is very interesting because it is able to
produce entangled states of up to tens of photons, relying on presently 
available laboratory equipment. In this paper we shall reconsider the 
scheme of Gheri {\it et al}. in order to make a detailed study of all 
the possible experimental limitations and sources of decoherence. 
We shall see that all the undesired decoherence effects can be
efficiently reduced using state-of-the-art technology, confirming 
therefore the preliminary results of Ref.~\cite{gheri}.

In Sec.~II we shall review the scheme of Ref.~\cite{gheri}; in 
Sec.~III we shall analyze its possible experimental
limitations, and we shall focus in particular on the effects of the intensity 
fluctuations of the driving lasers, of the various photon loss 
mechanisms, and of the motion of the atom trapped
within the cavity. Sec.~IV is for concluding remarks.

\section{The deterministic source of entangled photons of 
Gheri {\it et al.}}

The scheme consists,
in its simpest implementation, of an optical cavity containing
an atom (or ion) with a multi-level structure. As noticed in 
Ref.~\cite{gheri}, the same scheme could be however applied to a 
generic nonclassical medium in a superposition state replacing the 
atom (a quantum dot, for example). 
The cavity field is coupled through the mirrors to the continuum
of modes outside the resonator, which will sustain the desired
entangled single-photon wave packets. 
The relevant atomic level structure is a double three-level
$\Lambda $ scheme (see Fig.~1).
The levels $\vert i_{\alpha} \rangle$ and 
$\vert f_{\alpha} \rangle$ ($\alpha = 0,1$), may be, for example, 
hyperfine sublevels of the ground state, which are
coupled to the upper level $\vert
r_{\alpha} \rangle$ via, respectively, the classical field 
$\Omega_{\alpha}(t) e^{-i(
\omega_{\alpha} t + \phi_{\alpha}(t))}$, and the cavity mode with
annihilation operator $a_{\alpha}$, with
coupling constant $g_{\alpha}$. The index $\alpha$ referring to
the two distinct $\Lambda $ systems actually corresponds to the two
orthogonal polarizations of the cavity field. The lasers' 
amplitudes $\Omega_{\alpha}(t)$, their phases $\phi_{\alpha}$,
and their center frequencies $\omega_{\alpha}$ are therefore
the external control parameters. Both the laser fields and 
the two cavity modes are highly detuned ($\delta \gg \Omega_{\alpha}(t),
g_{\alpha}, \Gamma_{\alpha}$, where $\Gamma_{\alpha}$ are the 
spontaneous emission rates from $r_{\alpha}$) 
from the corresponding atomic transitions. Moreover 
their center frequency satisfies the condition $\delta \equiv 
\omega_{ri}^{\alpha}-\omega_{\alpha} = \omega_{rf}^{\alpha}-\omega_{c}$.
The large detuning serves the purpose of making the system insensitive
to the spontaneous emission from the excited levels $r_{\alpha}$, 
because in this case they are practically never populated. 
In such conditions, the excited levels 
can be adiabatically eliminated and the two $\Lambda $ systems becomes
equivalent to two effective two-level systems each coupled to a 
cavity mode with given polarization.
If we denote with $b_{\alpha}(\omega)$ 
the annihilation operator of the external electromagnetic field mode
with frequency $\omega$ and polarization $\alpha$, with the standard
bosonic commutation rule
\begin{equation} \label{eq:cm}
[b_{\alpha}(\omega),b^{\dagger}_{\beta}(\nu)] = \delta_{\alpha \beta}
\delta(\omega - \nu),
\end{equation}
the total Hamiltonian of the system in the interaction picture with 
respect to the free dynamics of the compound system, and after the 
adiabatic elimination of the $r$ levels, is
($\hbar =1$)
\begin{eqnarray} 
&&H = \sum_{\alpha} \frac{\Omega_{\alpha}^{2}(t)}{4\delta}
\vert i \rangle_{\alpha \alpha}
 \langle i \vert + \frac{g_{\alpha}^{2}}{\delta} a_{\alpha}^{\dagger}
a_{\alpha}\vert f \rangle_{\alpha \alpha} \langle f \vert+
i\frac{g_{\alpha}\Omega_{\alpha}(t)}
{2\delta} e^{i\phi_{\alpha}(t)}
a_{\alpha} \vert i \rangle_{\alpha \alpha} \langle f \vert \nonumber
\\
&&\label{eq:ham1}
-i\frac{g_{\alpha}\Omega_{\alpha}(t)}{2\delta} e^{-i\phi_{\alpha}(t)}
a_{\alpha}^{\dagger} \vert f \rangle_{\alpha \alpha}
\langle i \vert +
i \sqrt{\frac{k_{c\alpha}}{\pi}} \int d \omega
(a_{\alpha} b_{\alpha}^{\dagger}(\bar{\omega}) e ^{i \omega t} - 
(a_{\alpha}^{\dagger} b_{\alpha}(\bar{\omega}) e ^{-i \omega t}).
\end{eqnarray}
The first and second term are the ac Stark shifts due to the 
classical field and to the cavity field respectively; the third and 
the fourth term describe the Raman transition and the last term 
describes the interaction between the cavity modes
and the external continuum of modes, for which we have assumed, as 
usual \cite{qnoise}, a frequency-independent distribution
of coupling constants around the cavity mode frequency,
$\sqrt{k_{c\alpha}/\pi}$ ($k_{c\alpha}$ is the damping rate of the
cavity field with polarization $\alpha$
and $\bar{\omega} = \omega + \omega_{c}$).

Let us now see how the scheme works. The main idea is to transfer an 
initial coherent superposition of the atomic levels into a 
superposition of continuum excitations, by applying suitable laser 
pulses realizing the Raman transition between $i$ and $f$.
It is reasonable to assume 
that all the fields are initially in the vacuum state, so that
the initial state of the system is:
\begin {equation}\label{eq:stin}
\vert \psi (0) \rangle = \left( c_{0}\vert i \rangle_{0}+
c_{1}\vert i \rangle_{1}\right) \vert 0 \rangle_{c0} 
\vert 0 \rangle_{c1}\vert 0 \rangle_{r}
\end{equation}
(the subscript $r$ now denotes the state of the continuum).
There are no polarization mixing term and therefore the two branches 
corresponding to $\alpha=0$ and $\alpha=1$ will evolve independently. 
The state $\vert i \rangle_{\alpha}$ plays the role of the excited 
state and therefore, for each $\alpha$, the total excitation number
is one. The Hamiltonian (\ref{eq:ham1}) preserves the total excitation number 
and therefore the state at a generic time $t$ will be a coherent 
superposition of three terms, one in which the excitation is still
in the atom, one corresponding to the excitation stored in the cavity 
mode and one where the excitation has been transferred to the 
continuum of modes outside the cavity. The excitation transfer
is turned on and off by the laser pulses, and if
the laser pulse duration $T$ is much larger than the cavity decay time, 
$T\gg 1/k_{c\alpha}$, the amplitude corresponding to the excitation 
stored in the cavity mode will be completely negligible, so that one
has the state
\begin{equation}\label{eq:stf1}
\vert \psi_{1} \rangle = \vert 0 \rangle_{c0} 
\vert 0 \rangle_{c1} \otimes
\sum_{\alpha}c_{\alpha}\left[ e^{-\mu_{\alpha}(T)-
i\theta_{\alpha}(T)}\vert i \rangle_{\alpha} 
\vert 0 \rangle_{r}+ \vert f \rangle_{\alpha} \int d\omega G_{\alpha}(\omega,T)
b_{\alpha}^{\dagger}(\bar{\omega}) 
\vert 0 \rangle_{r}\right],
\end{equation}
where
\begin{eqnarray}\label{eq:1q}
\mu_{\alpha}(T) &=& \frac{g_{\alpha}^{2}}{4\delta^{2}k_{c\alpha}} \int_{0}^{T}
dt\Omega_{\alpha}^{2}(t) \\
\label{eq:2q}
\theta_{\alpha}(T)&=&\int_{0}^{T}dt\frac{\Omega_{\alpha}^{2}(t)}{4\delta}
\\
\label{eq:q3}
G_{\alpha}(\omega,T) &=& \sqrt{\frac{k_{c\alpha}}{\pi}} \int_{0}^{T}dt
e^{i\omega t}\frac{g_{\alpha}\Omega_{\alpha}(t)}{2\delta k_{c\alpha}}
e^{-\mu_{\alpha}(t)-i(\theta_{\alpha}(t) + \phi_{\alpha}(t))}.
\end{eqnarray}
The function $G_{\alpha}(\omega,T)$ determines the spectral envelope
of the single photon wave packet. An efficient transfer of the 
excitation to the external field is obtained when
$\mu_{\alpha}(T) \gg 1$ (which fixes a lower bound for the pulse area
of the applied laser field), and in this case the state of 
Eq.~(\ref{eq:stf1}) becomes 
\begin{equation}\label{eq:stf2}
\vert \psi_{1} \rangle = \vert 0 \rangle_{c0} 
\vert 0 \rangle_{c1} \otimes \sum_{\alpha}c_{\alpha}\int d \omega G_{\alpha}
(\omega,T)b_{\alpha}^{\dagger}(\bar{\omega})
\vert 0 \rangle_{r} \vert f \rangle_{\alpha}.
\end{equation}
This means that the generated wave packet is entangled with the atom. 
This entanglement can be transferred to a second photon which is 
subsequently created, by suitably recycling the system and performing 
a conditional measurement on the atomic levels. For the recycling
one has to apply first of all two $\pi$ pulses (one for each $\alpha$) in order to induce 
the transition from $|f\rangle _{\alpha}$ to $|i\rangle _{\alpha}$.
Moreover the continuum of modes outside the cavity has to be ``ready'' 
to receive a second, independent, photon wave packet. From an intuitive point 
of view, it is clear that if two successive wave packets do not
temporally overlap, i.e., the first wave packet is well far from the 
cavity when the second wave packet is generated, the two photons can 
be safely considered as individual qubits. From a formal point of 
view however this is not obvious since what one really has is just a 
continuum of e.m modes in a two-photon state. The independency of two 
successive photon wave packets can be however seen if 
we define the creation operator $B^{\dagger}(t_{j},T)$ of the wave-packet
generated during the time window $[t_{j},t_{j}+T]$ as
\begin{equation}\label{eq:wpco}
B_{\alpha}^{\dagger}(t_{j},T) = \int d \omega e^{i\omega t_{j}}
G_{\alpha}(\omega,T)b_{\alpha}^{\dagger}
(\bar{\omega})
\end{equation}
and notice that, when the two wave packets do not temporally overlap
($|t_{j}-t_{k}| \gg T$), the bosonic commutation relation 
\begin{equation}\label{eq:cmre}
[B_{\alpha}(t_{k},T),B_{\beta}^{\dagger}(t_{j},T)] \approx 
\delta_{\alpha \beta}\delta_{jk}
\end{equation}
holds. This allows us to regard each wave packet creation operator 
as acting on its own vacuum state and to define the following, 
independent, one 
photon states, with polarization $\alpha$ and generated in the
$j$-th generation sequence, $|\alpha 
\rangle_{j}=B_{\alpha}^{\dagger}(t_{j},T)|0\rangle_{r}$ ($t_{1}=0$).
In this way, the state after the first sequence of Eq.~(\ref{eq:stf2})
can be rewritten in the more compact form 
\begin{equation}
\vert \psi_{1}\rangle = 
\vert 0 \rangle_{c0} 
\vert 0 \rangle_{c1} \otimes \left(c_{0}\vert 0\rangle_{1}\vert f \rangle_{0}
+c_{1}\vert 1\rangle_{1}\vert f \rangle_{1}\right).
\end{equation}
If we recycle, i.e., $\vert f \rangle_{\alpha} \rightarrow 
\vert i \rangle_{\alpha}$, and restart the same procedure at the time
$t_{2} \gg t_{1}+T$, we get
\begin{equation}
\vert \psi_{2}\rangle = 
\vert 0 \rangle_{c0} 
\vert 0 \rangle_{c1} \otimes \left(c_{0}\vert 0\rangle_{1}
\vert 0\rangle_{2}\vert f \rangle_{0}
+c_{1}\vert 1\rangle_{1}\vert 1\rangle_{2}\vert f \rangle_{1}\right).
\end{equation}
More in general, the state of the system after $n$ 
generation cycles can be written as:
\begin{equation}\label{eq:stfn}
\vert \psi_{n} \rangle = \vert 0 \rangle_{c0} 
\vert 0 \rangle_{c1} \otimes \sum_{\alpha=0}^{1}
c_{\alpha} \vert \alpha
\rangle_{1}\vert \alpha \rangle_{2} \ldots \vert \alpha \rangle_{n}
\vert f \rangle_{\alpha}.
\end{equation}
The residual entanglement with the atom inside the cavity can 
eventually be broken up by making a measurement of the internal 
atomic state in an appropriate basis. For example, if one makes a 
projection measurement to etablish if the atom is in the 
$\left(\vert f \rangle_{0} +\vert f \rangle_{1}\right)/\sqrt{2}$ state
or in the $\left(\vert f \rangle_{0} -\vert f \rangle_{1}\right)/\sqrt{2}$ state,
the state of the e.m. continuum is correspondingly projected into the 
$n$-photons polarization-entangled state $c_{0} 
\vert 0 \rangle_{1}\vert 0 \rangle_{2} \ldots \vert 0 \rangle_{n}
\pm c_{1} \vert 1 \rangle_{1}\vert 1 \rangle_{2} \ldots \vert 1 
\rangle_{n}$.
This shows that many interesting entangled states, as the
four Bell states, the GHZ state, and its 
higher-dimensional generalization $\left(|\alpha_{1}, \alpha_{2},\ldots, 
\alpha_{n}\rangle + |1-\alpha_{1}, 1-\alpha_{2},\ldots, 
1-\alpha_{n}\rangle\right)/\sqrt{2}$ can be generated with this scheme.
As mentioned in \cite{gheri}, by coupling the levels 
$\vert f\rangle_{\alpha}$ with appropriate microwave pulses in between 
the generation sequence, one can also partially engineer the entanglement
and create a wider (even though not complete) set of $n$-photons 
entangled states.

\section{Possible decoherence sources}

In the preceding section we have seen how the scheme for the 
generation of polarization-entangled, time-separated, single photon 
wave packet works in the ideal case. In typical experimental 
situations there are however many physical mechanisms and technical
imperfections which may seriously limit the performance of the 
scheme. These are: (i) laser phase and amplitude fluctuations;
(ii) spontaneous emission from the excited levels $r_{\alpha}$
during the laser pulses; (iii) the effect of atomic motion; (iv)
random photon losses due to the absorption within the mirror or 
scattering; (v) systematic and random errors in the $\pi$ 
laser pulses used in the recycling process.

We have already seen that choosing a sufficiently large detuning 
$\delta$ of the optical frequencies from the atomic transitions 
connecting the excited levels $r_{\alpha}$ makes the scheme essentially
immune from the effect of atomic spontaneous emission (see also 
\cite{gheri}). Laser phase fluctuations also are not a problem because
the state produced after each cycle depends only on the {\em 
phase difference} between the two laser fields. This means that it is 
sufficient to derive the two beams from the same source, so that 
the fluctuations of the phase difference are essentially suppressed 
\cite{gheri}. The effect of imperfections in the recycling process
$\vert f\rangle _{\alpha} \rightarrow \vert i\rangle _{\alpha}$ is 
studied in detail in Ref.~\cite{gheri} assuming a random 
distribution of timings for the $\pi$ pulses and also the possibility 
of a dephasing angle between the two components. It is found that the 
process is robust against dephasing, while the correct timing of the 
$\pi$ pulses is a critical parameter.

Here we shall analyze in more detail the other three 
possible sources of decoherence, i.e., laser amplitude fluctuations, 
photon losses, and atomic motion, which have been discussed only 
briefly in Ref.~\cite{gheri}. In the next subsections we shall study 
the effect of these processes independently from each other.

\subsection{Intensity fluctuations of the laser pulses}

Amplitude fluctuations of the laser pulse mean fluctuations of the 
real parameters $\Omega_{\alpha}(t)$ which, as we can see from the 
Hamiltonian (\ref{eq:ham1}), in turn imply fluctuations of both the Stark 
shift term and the Raman coupling. Here we consider {\em intensity}
fluctuations, i.e., we assume that
the quantity $\Omega_{\alpha}^{2}(t)$ is the sum of a deterministic
signal and of a stochastic term, 
\begin{equation} \label{eq:wn}
\Omega_{\alpha}^{2}(t) = \Omega_{s\alpha}^{2}(t) +\sqrt{D_{\alpha}} \xi(t),
\end{equation}
where $\xi(t) $ is a zero-mean Gaussian white noise
with $\langle \xi(t) \xi(t') \rangle = \delta (t-t')$
and $D_{\alpha}$ is a diffusion coefficient quantifying
the strength of the fluctuations.

In the description of the ideal scheme of the preceding section, we 
have considered the optimal case $\mu_{\alpha}(T) \gg 1$, so that the 
probability of the initial excitation be still retained by the atom 
is zero. However, when we consider the repeated generation
of single-photon wave packets, we cannot neglect this small 
probability and the 
correct state of the system after the first cycle has to described by
Eq.~(\ref{eq:stf1}) rather than its ideal limit Eq.~(\ref{eq:stf2}).
By comparing the two states, it is evident that the fidelity of the 
preparation scheme, i.e., the probability to have the desired state
of Eq.~(\ref{eq:stf2}) after the first cycle is 
\begin{equation}\label{eq:fidf}
P(1) = \sum_{\alpha}\vert c_{\alpha}\vert^{2} \int d \omega \vert
G_{\alpha}(\omega,T) \vert ^{2}.
\end{equation}
However, using the normalization condition for the state 
(\ref{eq:stf1}), one has
\begin{equation}\label{eq:nc}
\int \vert G_{\alpha}(\omega,T) \vert ^{2} d \omega = 1 - 
e^{-2\mu_\alpha(T)},
\end{equation}
and iterating to the situation after $n$ cycles, one finds the 
fidelity for the generation of $n$ entangled photons
\begin{equation} \label{eq:fidf1}
P(n) = \sum_{\alpha}\vert c_{\alpha}\vert^{2}
\left[ 1 - e^{-2\mu_\alpha(T)}\right]^{n}.
\end{equation}
This expression shows that the laser intensity fluctuations
affect the efficiency of the process only through their effect on
the quantity $\mu_\alpha(T)$. This effect can be determined by
differentiating Eq.~(\ref{eq:1q}) and using Eq.~(\ref{eq:wn}); in 
this way one gets the following stochastic differential equation 
\begin{equation} \label{eq:sde1}
d\mu_{\alpha}(t) =\frac{g_{\alpha}^{2}}{4 \delta^{2} k_{c\alpha}}
[\Omega_{s\alpha}^{2}(t) dt +\sqrt{D_{\alpha}} dW(t))],
\end{equation}
where $W(t) \equiv \xi(t)dt$ is a Wiener process. This equation can 
be straighforwardly integrated so to get ($\mu_{\alpha}(0)=0)$
\begin{equation} \label{eq:sde2}
\mu_{\alpha}(T) =\frac{g_{\alpha}^{2}}{4 \delta^{2} k_{c\alpha}}
\left[\int_{0}^{T} \Omega_{s\alpha}^{2}(t)dt + \sqrt{D_{\alpha}} 
\int_{0}^{T}dW(t)\right].
\end{equation}
This shows that $\mu_{\alpha}(T)$ is a Gaussian stochastic variable, 
with mean value
\begin{equation}
\langle \mu_{\alpha}(T)\rangle = \frac{g_{\alpha}^{2}}{4 \delta^{2} k_{c\alpha}}
\int_{0}^{T} \Omega_{s\alpha}^{2}(t)dt,
\label{mean}
\end{equation}
and variance
\begin{equation}
\sigma^{2}_{\alpha}(T) = \langle \mu_{\alpha}(T)^{2}\rangle -
\langle \mu_{\alpha}(T)\rangle^{2} = 
\frac{g_{\alpha}^{4}D_{\alpha}T}{16 \delta^{4} k_{c\alpha}^{2}}.
\label{varia}
\end{equation}
In the presence of intensity fluctuations we have therefore to 
perform a Gaussian average of the fidelity of Eq.~(\ref{eq:fidf1}),
yielding
\begin{equation} \label{eq:prf}
\langle 1-e^{-2\mu_{\alpha}(T)}\rangle = 1-
\exp\left\{-2 \langle \mu_{\alpha}(T)\rangle -
\frac{g_{\alpha}^{4}}{8\delta^{4} k_{c\alpha}^{2}} D_{\alpha}T\right\};
\end{equation}
it is then reasonable to assume that the fluctuations in each pulse 
are independent, so that after the generation of $n$ wave packets,
the average value of
the fidelity is simply the product of terms of the form of Eq.~(\ref{eq:prf}).
Finally one has
\begin{equation} \label{eq:fdf}
\langle P(n) \rangle = \sum_{\alpha} \vert c_{\alpha} \vert ^{2}
\left(1-\exp\left\{-2 \langle \mu_{\alpha}(T)\rangle -
\frac{g_{\alpha}^{4}}{8\delta^{4} k_{c\alpha}^{2}} D_{\alpha}T\right\}\right)^{n}.
\end{equation}
In the case of square laser pulses with exact duration $T$ and 
intensity $I$, one has $\langle \mu_{\alpha}(T)\rangle =
g_{\alpha}^{2}IT/4\delta^{2}k_{c\alpha}$, and considering the simple case in 
which the parameters are the same for the two orthogonal polarizations,
the fidelity of preparation of the $n$-photons entangled state assumes 
the simple form
\begin{equation} \label{eq:fdf1}
\langle P(n)\rangle = \left(1-\exp\left\{-\frac{g^{2}IT}{2\delta^{2}k_{c}} -
\frac{g^{4}DT}{8\delta^{4}k_{c}^{2}}\right\}\right)^{n}.
\end{equation} 
Usually the intensity fluctuations are characterized in terms
of the relative error of the intensity pulse area, i.e.,
\begin{equation} \label{eq:fr}
F_{r} = \frac{\sqrt{DT}}
{\int_{0}^{T}\Omega_{s}^{2}(t)dt},
\end{equation}
which, in the case of the square pulse, becomes
$F_{r}=\sqrt{D/I^{2}T}$. The dependence of the fidelity upon the
laser intensity fluctuations is shown in Figs.~2 and 3. In Fig.~2 
$P(n)$ is shown as a function of the number of entangled photons $n$, 
for different values of $F_{r}$, while in Fig.~3
$P(n)$ is plotted versus $F_{r}$ for three different values of $n$
($n=3,5,10$). The case of a square pulse and 
identical parameters for the two polarizations is considered
(see the captions for parameter values). These figures, and Fig.~3 in 
particular, show that laser intensity fluctuations do not significantly 
affect the performance of the scheme, even at quite large relative 
fluctuations.

\subsection{Photon losses}

An important source of errors is the fact that, in the generation 
scheme of the output wave packet, the photon can be lost. To be 
specific, the photon can be absorbed by the cavity mirrors, or it can 
be transferred to ``undesired'' external electromagnetic modes of the
continuum, different from the monitored, output modes. This may happen due  
to scattering losses, or due to the transmission through the other 
(imperfect) cavity mirror, 
different from the output coupling mirror. It is quite reasonable
to assume that both the ``undesired'' electromagnetic modes of the 
continuum outside the cavity, and the internal degrees of freedom 
of the mirrors which can be excited by 
the cavity mode, can be represented by a continuum of
harmonic oscillators with annihilation operator 
$m_{\alpha}(\omega)$, satisfying the usual bosonic commutation relations
$\left[m_{\alpha}(\omega), m_{\beta}^{\dagger}(\nu)\right] = 
\delta_{\alpha \beta} \delta(\omega - \nu)$, as it is done for the 
monitored, output electromagnetic modes, described 
by the bosonic operators $b_{\alpha}(\omega)$.
It is also reasonable to assume that the coupling between 
the two cavity modes and the ``environmental''  modes
can be described in the same way as in Eq.~(\ref{eq:ham1}), so that
the total Hamiltonian of the system, in the interaction picture with 
respect to the free Hamiltonian of the compound system (which now 
includes also these environmental modes), becomes
\begin{equation}\label{eq:ham2}
H_{tot} = H + \sum_{\alpha} i \sqrt{\frac{k_{a\alpha}}{\pi}} \int d \omega
(a_{\alpha} m_{\alpha}^{\dagger}(\bar{\omega}) e^{i \omega t} -
a_{\alpha}^{\dagger} m_{\alpha}(\bar{\omega}) e^{-i\omega t}),
\end{equation}
where $H$ is the Hamiltonian of Eq.~(\ref{eq:ham1}). 
As in Eq.~(\ref{eq:ham1}), we have considered a frequency-independent
(but polarization-dependent)
coupling constant $\sqrt {k_{a\alpha}/\pi}$ centered around 
the cavity mode frequency $\omega_{c}$, where $k_{a\alpha}$ is the 
decay rate into the undesired modes for the photons with polarization 
$\alpha$. 

At this point one could generalize the calculations already developed in 
\cite{gheri} for the model Hamiltonian (\ref{eq:ham1}) 
to the more general Hamiltonian of Eq.~(\ref{eq:ham2}),
and derive the fidelity of the $n$ entangled photons generation process
in the presence of photon losses. However, one can easily understand that the 
results of Ref.~\cite{gheri} can be immediately adapted to the present 
case, thanks to the simplicity of the above modelization of the 
various absorption processes. In fact, the interaction term added in 
Eq.~(\ref{eq:ham2}) implies adding a {\em supplementary decay channel} for 
the cavity photons, in addition to the standard channel provided by 
the output mirror. Since, for a cavity photon with polarization 
$\alpha$, $k_{c\alpha}$ is the probability to be transmitted to the 
desired output e.m. modes per unit time, 
and $k_{a\alpha}$ is the loss probability per unit time, it is evident 
that the probability to produce the correct single photon wave packet
of Eq.~(\ref {eq:stf2}) in a given cycle has to be corrected by the 
factor $k_{c\alpha}/\left(k_{c\alpha}+k_{a\alpha}\right)$ for each 
$\alpha$. As we have seen in the preceding 
section, two successive single-photon wave 
packets have to be well separated in time (and therefore in space) in 
order to be safely considered as independent qubits,
and therefore it is reasonable 
to assume that the eventual photon absorption processes taking place
in two generation cycles are independent. This implies that in the 
general case of the generation of an $n$-photons entangled state, the 
correction factor to the fidelity $P(n)$ due to the photon losses is
$\left[k_{c\alpha}/\left(k_{c\alpha}+k_{a\alpha}\right)\right]^{n}$.
Therefore, using the general expression (\ref{eq:fidf1}) for the 
fidelity, and taking into account that in the presence of photon 
absorption, the decay rate $k_{c\alpha}$ has to be replaced by
the total photon decay rate $k_{c\alpha}+k_{a\alpha}$
in the expression (\ref{eq:1q}) for $\mu_{\alpha}(T)$, one finally 
arrives at the following expression of the fidelity of generation
of an $n$-photons entangled state in the presence of photon losses
\begin{equation}\label{eq:fidag2}
P(n) = \sum_{\alpha} \vert c_{\alpha} \vert^{2}
\left( \frac{k_{c\alpha}}{k_{c\alpha} + k_{a \alpha}} \right) ^{n}
\left(1-\exp\left\{-\frac{g_{\alpha}^{2}}{2 \delta^{2}(k_{c\alpha}+
k_{a \alpha})}
\int_{0}^{T} dt \Omega_{\alpha}^{2}(t)\right\}\right)^{n}
\end{equation}
In the case of square pulses with intensity $I$ and duration $T$,
and identical parameters for both polarizations, the above equation
becomes
\begin{equation}\label{eq:fdarp}
P(n) = \left( \frac{k_{c}}{k_{c} + k_{a}} \right) ^{n}
\left(1-\exp\left\{-\frac{g^{2}IT}{2 \delta^{2}(k_{c}+k_{a})}\right\}\right)^{n}.
\end{equation}
The behaviour of the fidelity of preparation of $n$ entangled photons
in the presence of photon losses is shown in Figs.~4 and 5: in Fig.~4
$P(n)$ is plotted versus $n$ for different values of the ratio
$k_{a}/k_{c}$, while in Fig.~5 $P(n)$ is plotted as a function of 
$k_{a}/k_{c}$, for $n=3,5,10$ (from the upper to the lower curve).
The case of square laser pulses and identical parameter for the two 
polarizations is again considered. 
One can see that photon losses, differently from laser intensity 
fluctuations, can seriously limit the performance of the
scheme and that the fidelity of preparation rapidly decays for 
increasing photon losses. 

In Ref.~\cite{gheri} the limitations due to photon losses are briefly 
discussed and it is proposed that they can be avoided using 
postselection schemes, that is, detecting the photons and discarding 
the cases corresponding to a number of detected photons smaller than 
$n$. In this way, however, the scheme ceases to be a deterministic 
source, able to produce entangled photons on demand. It becomes 
instead a conditional source, in which the entangled photons are no 
more available after detection, and in which the quality of the state 
is established only {\em a posteriori}. The fidelity of 
Eqs.~(\ref{eq:fidag2}) and (\ref{eq:fdarp}) instead refers to the more 
general case in which no conditional measurements are made, and all the 
events are considered. In such a case the proposed source remains
a deterministic source of entangled photons even in the presence
of photon losses, even though with a lower fidelity.

\subsection{Atomic motion}

Up to now we have assumed the atom to be in a fixed position within 
the cavity. However, the atomic center-of-mass motion may affect the 
performance of the scheme, by inducing fluctuations and dephasing of 
the internal atomic states. To state it in an equivalent way, the 
atomic motional degrees of freedom will generally 
get entangled with the internal ones and, in turn, with the cavity 
modes, and this may lead to decoherence and quantum information loss.

It is evident that the optimal way to minimize the effect of atomic 
motion is to trap and cool the atom, possibly to the motional ground 
state of the trapping potential. Cooling to the motional ground state has been
already achieved both with ions in rf-traps \cite{ion}, and with 
neutral atoms in optical lattices \cite{neutr} using resolved sideband 
cooling, which requires operating in 
the Lamb-Dicke regime, where the size of the atomic wave packet $L$ 
is much smaller than the optical wavelength $\lambda =2\pi 
c/\omega_{c}$, and strong confinement. 
The effect of spatial variation is minimized if the 
minimum of the trapping potential coincides with an antinode of both 
the cavity mode and of the laser field. This implies that also the 
two classical lasers have to be in a standing wave configuration.

Therefore we shall assume that the atom is trapped in 
some way (ion in a rf-trap, or neutral atom in a far off resonance 
dipole trap) in a harmonic potential with frequency $\omega_{0}$,
near an antinode of the cavity
field, which we choose as the origin for the spatial coordinates.
Taking into account the spatial dependence of both the laser
and the cavity field, and considering for simplicity only the 
one-dimensional motion along the cavity axis $\hat{x}$,
the Hamiltonian of the system becomes
\begin{eqnarray}
&&H'=\frac{\hat{p}^{2}}{2m}+ \frac{1}{2} m \omega_{0}^{2}\hat{x}^{2} +
\sum_{\alpha} \frac{\Omega_{\alpha}^{2}(t)}{4\delta}
\cos^{2}k_{L}\hat{x}\vert i \rangle_{\alpha \alpha}
 \langle i \vert + \frac{g_{\alpha}^{2}}{\delta} \cos^{2}k_{r}\hat{x} 
 a_{\alpha}^{\dagger}
a_{\alpha}\vert f \rangle_{\alpha \alpha} \langle f \vert \nonumber
\\
&&+\label{eq:hamat}
i\frac{g_{\alpha}\Omega_{\alpha}(t)}
{2\delta} \cos k_{L}\hat{x} \cos k_{r}\hat{x} \left(e^{i\phi_{\alpha}(t)}
a_{\alpha} \vert i \rangle_{\alpha \alpha} \langle f \vert 
- e^{-i\phi_{\alpha}(t)}
a_{\alpha}^{\dagger} \vert f \rangle_{\alpha \alpha}
\langle i \vert\right) +
i \sqrt{\frac{k_{c\alpha}}{\pi}} \int d \omega
(a_{\alpha} b_{\alpha}^{\dagger}(\bar{\omega}) e ^{i \omega t} - 
(a_{\alpha}^{\dagger} b_{\alpha}(\bar{\omega}) e ^{-i \omega t}),
\end{eqnarray}
where $k_{L}$ and $k_{r}$ are the laser field and cavity mode field
wave vector respectively, $m$ is the
mass of the atom and $\hat{p}$ its momentum.
As discussed above, optimal conditions for the generation scheme
are obtained in the Lamb-Dicke 
limit, which implies approximating the cosine terms in the 
Hamiltonian (\ref{eq:hamat}) at second order, i.e., 
\begin{eqnarray}
&&\cos^{2}k_{L}\hat{x} \simeq 
1-\eta_{L}^{2}\left(l+l^{\dagger}\right)^{2} \label{ld1} \\
&&\cos^{2}k_{r}\hat{x} \simeq 
1-\eta_{r}^{2}\left(l+l^{\dagger}\right)^{2} \label{ld2} \\
&&\cos k_{L}\hat{x} \cos k_{r}\hat{x}\simeq 
1-\left(\frac{\eta_{L}^{2}+\eta_{r}^{2}}{2}\right)
\left(l+l^{\dagger}\right)^{2}, \label{ld3} 
\end{eqnarray}
where $\eta_{j}=k_{j}\sqrt{\hbar/2m\omega_{0}}$ ($j=L,r$), are the two 
Lamb-Dicke parameters and $l$ is the annihilation operator for the 
vibrational quanta.

In general, besides the Hamiltonian evolution driven by (\ref{eq:hamat}), the 
atomic center-of-mass motion is also affected by non-unitary processes 
such as the cooling, the recoil due to the spontaneous emission, and 
heating processes caused by technical imperfections such as fluctuating 
electric potentials in the trapped ion case \cite{nist},
or intensity fluctuations of the laser used in the case of optical dipole traps
\cite{ohara}. The atomic recoil is negligible in the Lamb-Dicke limit;
moreover it is in principle possible to turn the laser cooling on 
whenever needed, and therefore in this case it is reasonable to
neglect the heating processes.
This means that the 
atomic vibrational motion can be satisfactorily described by the 
Hamiltonian (\ref{eq:hamat}) (supplemented with 
(\ref{ld1})-(\ref{ld3})). However, it is realistic to assume that the 
cooling process will not be perfect and 
leave some residual vibrational excitation, which 
can be described as an effective thermal state $\rho_{N}^{vib}$
with mean vibrational number
$N$ for the initial state of the atomic 
center-of-mass motion at every generation cycle.
Therefore, the state of the whole system at the beginning of a cycle
will be
\begin{equation}
\rho_{tot}(0)=\vert \psi(0)\rangle \langle \psi(0)\vert \otimes 
\rho_{N}^{vib}, \label{inivib}
\end{equation} 
where $\vert \psi(0)\rangle $ is given by Eq.~(\ref{eq:stin}).
The probability to generate the right entangled state after the first 
cycle is 
\begin{equation}
P(1)={\rm Tr}_{vib}\left\{\langle \psi_{1}\vert \rho_{tot}(T)\vert 
\psi_{1}\rangle \right\}, \label{fidevib}
\end{equation}
where $\vert \psi_{1}\rangle $ is the desired state to generate of 
Eq.~(\ref{eq:stf2}), 
$\rho_{tot}(T)$ is the state of the total system (including the atomic 
center-of-mass) at the end of the cycle, and ${\rm Tr}_{vib}$ denotes 
the trace over the vibrational degree of freedom.
This fidelity after the first cycle has been calculated numerically 
using the Hamiltonian of Eq.~(\ref{eq:hamat}) (in the Lamb-Dicke limit)
and the initial condition (\ref{inivib}). This calculation is 
simplified by the fact that the 
excitation number operator for a given polarization,
\begin{equation}\label{eq:exnumop}
{\cal N} = \vert i \rangle \langle i \vert + a^{\dagger} a  +
 \int d \bar{\omega} b^{\dagger}(\bar{\omega})b(\bar{\omega})
\end{equation}
is a constant of motion even when the atomic motion is considered.
Since the initial excitation number is ${\cal N}=1$, the evolution 
will always be confined within the subspace with only one excitation. 
In the general case of $n$ generation cycles, the 
temporal separation of two successive wave packets guarantees that
the preparation fidelity of $n$ entangled photons will be simply the 
$n$-th power of $P(1)$,
\begin{equation}
P(n)=\left[{\rm Tr}_{vib}\left\{\langle \psi_{1}\vert \rho_{tot}(T)\vert 
\psi_{1}\rangle \right\}\right]^{n}. \label{fidevib2}
\end{equation} 
The numerical results for the fidelity $P(n)$ are plotted 
as a function of the number of entangled photons in Fig.~6, for 
increasing values of the initial effective mean vibrational number $N$
(from the upper to the lower curve). One can see that if the residual 
vibrational excitation left by the cooling process is appreciable
($N \simeq 1$, the lower curve of Fig.~6), the fidelity of preparation 
can be seriously affected, while the effect of atomic motion is modest 
when the cooling process is efficient ($N < 0.1$).

\section{Conclusions}

In this paper we have studied the sensitivity to the various possible 
sources of decoherence of a recently proposed 
scheme \cite{gheri} for the deterministic generation of polarization-entangled 
single photon wave packets. The scheme employs a trapped and 
laser-cooled atom within a cavity in a double three level $\Lambda$ 
scheme. The successively generated single-photon wave packets remain 
entangled with the atom and an appropriate conditional measurement on 
the atomic internal levels transfers the entanglement to the set of 
photons. These wave packets can be considered as independent qubits 
as long as they are well separated in time. 
The scheme of Ref.~\cite{gheri} can be particularly suited for the 
implementation of recently proposed multi-party quantum communication 
schemes based on quantum information sharing \cite{hillery,bose}.
Here we have focused in 
particular on the limiting effects which may be caused by laser intensity 
fluctuations, photon losses, and by the atomic motion, which have been 
discussed only briefly in \cite{gheri}. 
Photon losses prove to be the predominant limiting factor, while the 
scheme is robust against the effect of laser intensity fluctuations. 
Atomic motion does not seriously limit the performance of the scheme, 
but only if the atom is sufficiently cooled close to the ground state
of the trapping potential, otherwise the residual vibrational excitation 
can significantly lower the fidelity of preparation.

\section{Acknowledgments}

This work has been partially supported by INFM through 
the PAIS ``Entanglement and decoherence''.

\begin{figure}
\centerline{\epsfig{figure=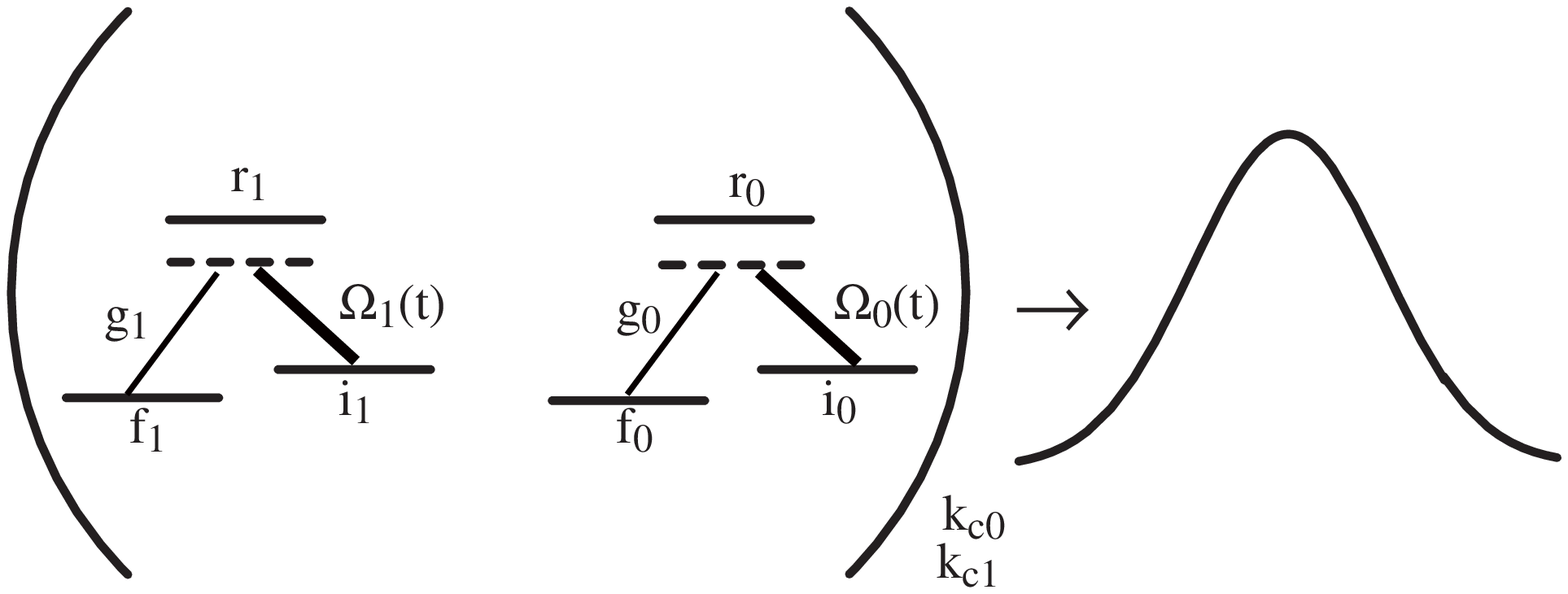,width=10cm}}
\caption{Schematic description of the cavity QED scheme
proposed by Gheri {\it et al}. for the generation of 
polarization-entangled photon wave packets. The cavity and the 
relevant level structure of the atom trapped in it are shown.}
\label{fidapp}
\end{figure} 

\begin{figure}
\centerline{\epsfig{figure=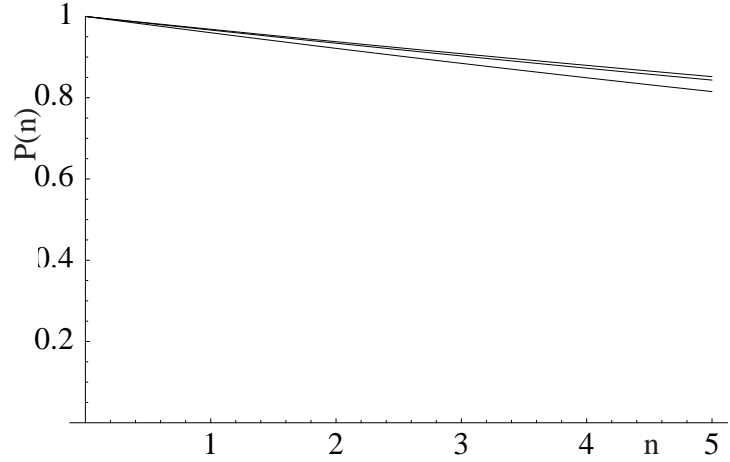,width=10cm}}
\caption{Fidelity of preparation $P(n)$ versus the number of entangled
photons $n$ for different values of the relative
fluctuations of the laser intensity $F_r$ (see Eq.~(\protect\ref{eq:fr})).
The three curves refer, starting from the upper one, 
to $F_r =0,0.1,0.2$. Square laser pulses and identical parameters for 
the two polarizations have been considered.
The other parameter values are: $g=\sqrt{I} =
60$ MHz; $\delta =1500$ MHz; $k_c = 25$ MHz; $T= 30$ $\mu$sec.}
\label{fidlas}
\end{figure} 

\begin{figure}
\centerline{\epsfig{figure=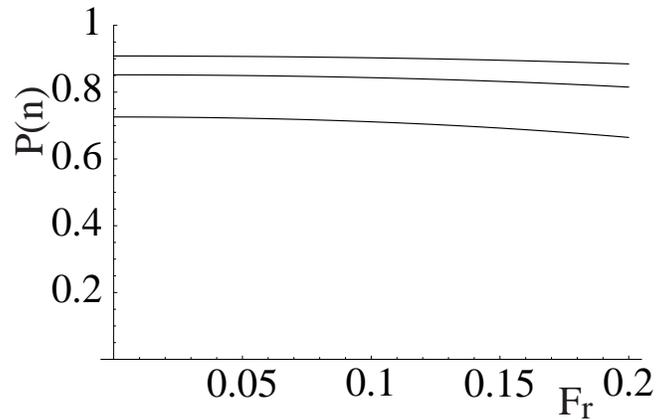,width=9cm}}
\caption{Fidelity of preparation $P(n)$ versus the 
the relative fluctuations of the laser intensity $F_r$ 
for different values of the number of generated entangled wavepackets
$n$. 
The three curves refer, starting from the upper one, 
to $n=3,5,10$. Square laser pulses and identical parameters for 
the two polarizations have been considered.
The other parameter values are the same as those of 
Fig.~\protect\ref{fidlas}.}
\label{fidnlas}
\end{figure} 

\begin{figure}
\centerline{\epsfig{figure=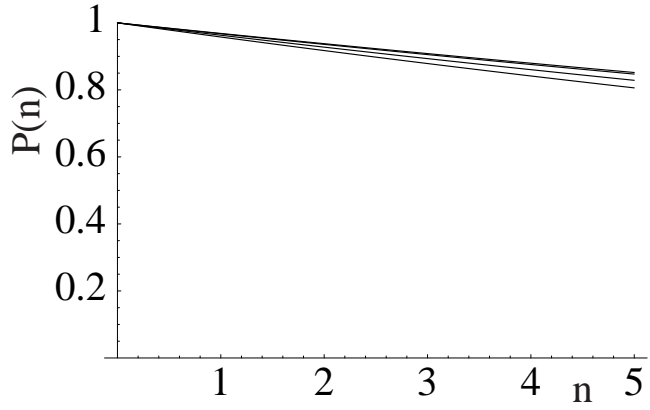,width=9cm}}
\caption{Fidelity of preparation $P(n)$ versus the number 
of photon wavepackets $n$ in the case of no laser intensity 
fluctuations but in the presence of photon losses with a rate
$k_{a}$. 
The four curves refer to different values of the ratio 
of decay rates $k_a/k_c$: starting from the upper one, 
$k_a/k_c = 0, 0.001, 0.005, 0.01$. 
Square laser pulses and identical parameters for 
the two polarizations have been considered.
The other parameter values are as in Fig.~\protect\ref{fidlas}.}
\label{fidass}
\end{figure} 

\begin{figure}
\centerline{\epsfig{figure=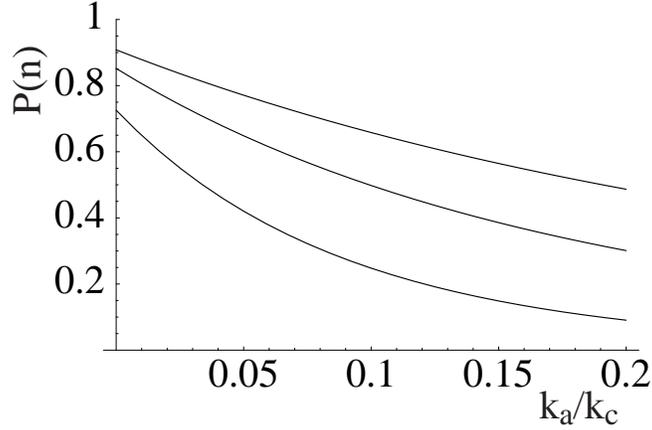,width=9cm}}
\caption{Fidelity of preparation $P(n)$ versus the ratio of decay rates
$k_a/k_c$
for three different number of generated entangled wavepackets $n$. 
The three curves refer, starting from the upper one, 
to $n=3,5,10$. Square laser pulses and identical parameters for 
the two polarizations have been considered.
The other parameter values are the same as those of 
Fig.~\protect\ref{fidlas}.}
\label{fidnass}
\end{figure}

\begin{figure}
\centerline{\epsfig{figure=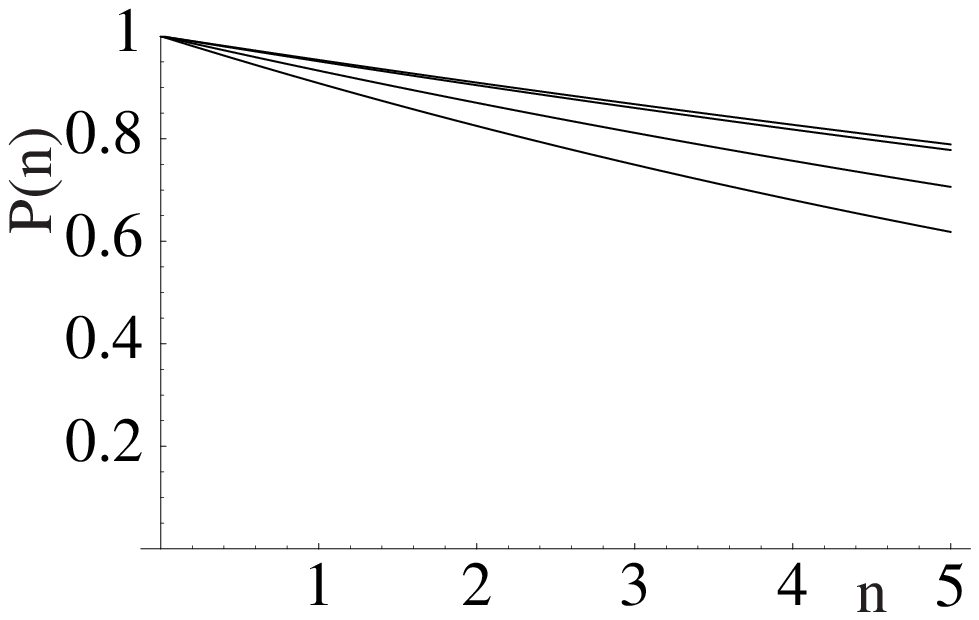,width=9cm}}
\caption{Fidelity of preparation $P(n)$ versus the number 
of photon wavepackets $n$ for different values of the
effective initial mean vibrational thermal number $N$.
The four curves refer, starting from the upper one, 
to $N=0.01, 0.1, 0.5, 1$. The other parameters for the atomic motion 
are $\omega_0 = 1$ MHz and $\eta_r = \eta_l = 0.07$, while
the other parameter values are the same as those of 
Fig.~\protect\ref{fidlas}. Square laser pulses and identical parameters for 
the two polarizations have been considered.}
\label{fidmoto}
\end{figure} 


\begin{references}
\bibitem{cryp}A. Muller, H. Zbinden, and N. Gisin, 
Europhys. Lett. 33, 335 (1996);
W.T. Buttler et al., Phys. Rev. Lett. 81, 3283 (1998).
\bibitem{telep} D.~Bouwmeester, J.-V.~Pan, K.~Mattle, M.~Eibl, H.~Weinfurter,
and A.~Zeilinger, Nature (London) {\bf 390}, 575 (1997);
D.~Boschi, S.~Branca, F.~De~Martini, L.~Hardy, 
and S.~Popescu, Phys. Rev. Lett. {\bf 80}, 1121 (1998);
A. Furusawa et al., Science {\bf 282}, 706 (1998).
\bibitem{kwiat}P.~G.~Kwiat, K.~Mattle, H.~Weinfurter, A.~Zeilinger,
A.~V.~Sergienko, Y.~H.~Shih, Phys. Rev. Lett. {\bf 75}, 4337 (1995);
P.~G.~Kwiat, E.~Waks, A.~G.~White, I.~Appelbaum, and
P.~H.~Eberhard, Phys. Rev. A {\bf 60}, R773 (1999).
\bibitem{ghz}D. M. Greenberger, M. A. Horne, and A. Zeilinger, 
Phys. Today {\bf 46}, No. 8, 22 (1993).
\bibitem{ghz2}D. Bouwmeester, J-W. Pan, M. Daniell, 
H. Weinfurter, and A. Zeilinger, Phys. Rev. Lett. {\bf 82}, 1345 
(1999).  
\bibitem{law}C. K. Law and H. J. Kimble, J. Mod. Opt. {\bf 44}, 2067 (1997).
\bibitem{yama}A. Imamo\u{g}lu and Y. Yamamoto, Phys. Rev. Lett. {\bf 
72}, 210 (1994).
\bibitem{gheri}K. M. Gheri, C. Saavedra, P. T\"orma, J. I. Cirac, and 
P. Zoller, Phys. Rev. A {\bf 58}, R2627 (1998).
\bibitem{qnoise}C. W. Gardiner, {\it Quantum Noise} (Springer, Berlin, 
1991).
\bibitem{ion}F. Diedrich {\it et al}., 
Phys. Rev. Lett. {\bf 62}, 403 (1989); C. Monroe {\it et al}.,
 Phys. Rev. Lett. {\bf 75}, 4011 
(1995); E. Peik {\it et al}., Phys. Rev. A {\bf 60}, 439 (1999); 
Ch. Roos {\it et al}.,
Phys. Rev. Lett. {\bf 83}, 4713 (1999).
\bibitem{neutr} S.E. Hamann {\it et al}., Phys. Rev. Lett. {\bf 80}, 
4149 (1998);
H. Perrin {\it et al}., Europhys. Lett. {\bf 42}, 95 (1998).
\bibitem{nist}
D.J. Wineland, C. Monroe, W.M. Itano, 
D. Leibfried, B.E. King,
D.M. Meekhof, J. Res. Natl. Inst. Stand. Technol. {\bf 103}, 259 (1998).
\bibitem{ohara}
T.A. Savard, K.M. O'Hara, and J.E. Thomas, Phys. Rev. A {\bf 56},
R1095 (1997).
\bibitem{hillery}M. Hillery, V. Bu\v{z}ek, and A. Berthiaume, Phys. Rev. A {\bf 
59},1829 (1999).
\bibitem{bose}S. Bose, V. Vedral, and P. L. Knight, Phys. Rev. A {\bf 
57}, 822 (1998).




\end{references}
\end{document}